# Real Time Global Tests of the ALICE High Level Trigger Data Transport Framework

B. Becker, S. Chattopadhyay, C. Cicalò J. Cleymans, G. de Vaux, R .W. Fearick, V. Lindenstruth, M. Richter, D. Rörich, F. Staley, T.M. Steinbeck, A. Szostak, H. Tilsner, R. Weis and Z.Z. Vilakazi.

***Abstract*—The High Level Trigger (HLT) system of the ALICE experiment is an online event filter and trigger system designed for input bandwidths of up to 25 GB/s at event rates of up to 1 kHz. The system is designed as a scalable PC cluster, implementing several hundred nodes. The transport of data in the system is handled by an object-oriented data flow framework operating on the basis of the publisher-subscriber principle, being designed fully pipelined with lowest processing overhead and communication latency in the cluster. In this paper, we report the latest measurements where this framework has been operated on five different sites over a global north-south link extending more than 10,000 km, processing a "real-time" data flow.**

## I. Introduction

For the future Large Hadron Collider (LHC) [1], [2] at CERN, four new experiments ALICE [3], ATLAS [4], CMS [5] and LHCb [6] are currently being built. They have the principal tasks of searching for the Higgs-Boson and physics beyond the standard model (ATLAS and CMS), CP violation (LHCb), and heavy-ion physics with a particular emphasis on quark-gluon plasma research [7]. Each of these experiments will produce large data sets that have to be handled by their trigger and data acquisition systems. In all four experiments large PC farms will handle a significant part of the generated data flow by performing event trigger, filter, and selection tasks. For the ALICE experiment these tasks are the primary function of the High Level Trigger (HLT) system [8], which is designed to implement up to 1000 multiprocessor 19" commercial PC-type computers. All three functions of the HLT require reconstruction and pattern recognition to be performed online, analysing data streams of 25 GB/s under real-time conditions.

B. Becker and C. Cicalò are with the I.N.F.N, Sezione di Cagliari, Cittadella Universitaria di Monseratto, Casella Postale 170 09042 Monseratto (Cagliari), Italy.

S. Chattopadhyay is with the Saha Institute, 1/AF Bidhannagar, Kolkata 700 064, India

J. Cleymans, G. de Vaux, R. W. Fearick, A. Szostak and Z. Z. Vilakazi are with the Department of Physics, University of Cape Town, Private Bag Rondebosch, 7000, South Africa

V. Lindenstruth, T.M. Steinbeck, H. Tilsner and R. Weis are with the Kirchoff Institute, Ruprecht-Karls University Heidelberg, D-69120 Heidelberg, Germany

M. Richter and D. Rörich are with the Department of Physics, University of Bergen, Allegaten 55, N-5007 Bergen, Norway

F. Staley is with the DAPNIA/SPhN CEA Saclay, Fr-91191 Gif sur Yvette Cedex, France

This work is supported by the National Research Foundation (NRF2357), South Africa; INFN-Cagliari, Italy; Commissariat à l'Energie Atomique, Saclay, France; BMBF (06HD157I), Germany.

For the transport of the data inside the HLT system a software framework has been developed based upon the publisher-subscriber principle [9], also known as the producer-consumer paradigm. The publisher-subscriber design is particularly suited to distributed data-driven systems where flexibility is required by, for example, the need to specify a configuration only at run time. The implementation of this design in the ALICE HLT framework has already been used in a large number of tests and benchmarks as well as in beam test scenarios of detector components. Through its use of simple components, plugged together via well-defined interfaces, it provides a flexible and easily reconfigurable capability for the controlling the data flow. For efficiency reasons data copying is kept to a minimum inside a node. This is achieved by placing the data into shared memories and exchanging only descriptors to that data via named pipes. Dedicated software exist to connect components on different nodes. After the final selection and compression of the data inside the HLT it is recorded to permanent storage for later collaborative off-line detailed analysis. The event selection by the HLT serves to reduce the overall rate of data written to permanent storage

One characteristic feature of each of the four mentioned experiments is that they consist of large globally distributed collaborations, with several tens to more than 150 institutes and more than 1000 scientists taking part. The HLT provides an real-time filter for the data acquisition system by selecting which events to write to permanent storage, thereby achieving a background suppression of several factors, depending on the particular signal and background processes (see [10] for a discussion of the data suppression capabilities of the HLT). Nonetheless, the set of data acquired during a year-long run of the experiment will exceed 1 PB. For the task of analysing these large data sets which were preselected by the on-line HLT system, a global distributed effort is necessary, not least because more than 10 000 processors are required to accomplish this task. Each event has a size of 12.5 MB independent from the next, implying that parallel computing can be exploited to a very large degree in the experiment's computing model (see reference [11]). Therefore off-line processing is planned to be performed on a computing grid - mostly at sites participating in the LHC Computing Grid, LCG [12] - exploiting the independence of the individual events.

The on-line processing systems, having real-time characteristics, are typically run on the computer farm close to the experiment. The algorithms implemented in the ALICE HLT typically have the characteristic that they operate on sub-event level data, exploiting to the largest degree possible parallelism in the data to increase performance. The real-time, data rate and most importantly the reliability constraints of the on-line systems make it prohibitive to currently operate them as grid application. In particular it is desirable for such a critical part of the experimental apparatus as the trigger to be as close as possible to the autonomous experimental hardware, in order to ensure the highest possible availability.

However, if the on-line communication architecture is designed carefully, avoiding round trip latencies in the communication and flow control paths it may be possible to perform even on-line trigger and filter functionality in a widely distributed fashion, such as a grid-like system. Besides the trigger processing, it is also possible that the same data transport framework could be used to perform similar, but less critical processing tasks, such as off-line processing. At this point the presented data driven, distributed grid processing infrastructure is being studied amongst others with respect to its application to the distributed radio astronomy system LOFAR, where multiple, distributed sites produce data streams, which have to be pre-processed individually and then combined with other sites. This process bears a lot of similarities to the ALICE HLT architecture. The following article details the first successful test of the ALICE HLT on-line trigger system over distances exceeding 10000 km in a proof-of-principle test. We begin with an overview of the software design, giving a description o the framework, its components and how configuration is managed. The actual test configuration and results are given in section III, and we give our conclusions and a summary in section IV.

## II. Overview of the Software

### A. Data Transport Framework Design

The system architecture is based on the publisher-subscriber paradigm because this allows a dynamically determined number of data consumers to connect to the appropriate producer. This design allows dynamic reconfiguration and inherent fault tolerance as detailed in the following sections. Another added consequence of this design choice is encapsulation of the actual processing code, into individual modules, connected by the publish-subscriber interface. In the design of the particular publisher/subscriber interface, particular emphasis was given to three important points: efficiency, flexibility, and fault tolerance. For a full description of this functionality of the framework, we refer the reader to [13].

Efficiency is important for the communication framework because the analysis of the event data is very compute intense. This is achieved by placing the data into shared memory segments by its publishing object and by passing descriptors to the subscribers via named pipes. When all subscribers have subsequently informed the publisher that they have finished processing a given event, it is released and the shared memory region can be re-used.

Flexibility has to be present in the framework as the configuration has to adopt to dynamically changing requirements of the experiment and the analysis. The primary mechanism for providing flexibility is the separation of the framework into components which can be dynamically connected in different configurations. Using the data flow components defined below, any processing hierarchy can be constructed. As the publisher-subscriber supports dynamic connections and disconnections at runtime, the system configuration can be adapted while it is active.

Fault tolerance is also achieved, using the dynamic reconfiguration capability of the communication building blocks. For instance, it allows for the replacement of failed components during runtime and also for the addition and/or removal of components in the data stream as required in reaction to dynamic events occurring in the system. This feature is in particular important in a distributed grid-like environment, where public networks are being used and many operational conditions are out of the control of the operators. A second major fault tolerance building block is related to the bridge components connecting different nodes as described in more detail below. These components also have the ability to establish connections dynamically at runtime, not only for re-establishing existing connections but also for establishing new connections between nodes. This mechanism allows the isolation of faulty nodes in the system and to replace them with stand-by nodes. In essence the HLT fault tolerance architecture uses both a bottom-up and a top down approach at certain levels. The bottom-up aspect ensures that all modules of the system are somewhat independent and capable of dynamic reconfiguration. The top-down aspects of the framework's functionality implement the intelligence to discover and react to any issues in the system, in a semi-automated way, and issue the appropriate commands to the fundamental HLT communication framework. The actual analysis code is not affected and completely independent.

### B. Framework Components

The HLT communication framework components can be categorized into three groups:

1) **Dataflow components** are utility programs designed to shape the dataflow in a framework system.
2) **Application components** perform the actual on-line data processing and encapsulate the analysis code between a subscriber and publisher object. There is also a number of maintenance application programs, such as data integrity and performance checkers, dummy routines, etc.
3) **Application component templates** provide a base from which application components may be constructed for a specific system.

Application components templates and application components exist in three variations:

- Data input source components are the points where data is inserted into a dataflow chain constructed with the framework. They access entities external to the framework and make their data available to other framework components.
- Data processing components perform the work inside a framework system. They accept data from other components, process it to produce new output data, and make this new data available again to other framework components. By chaining multiple processing components together complex analysis processes can be performed.
- Data sink components act as output of a dataflow chain. They accept data from other framework components and can transmit this data to entities outside of the framework.

There are 5 primary dataflow components contained in the framework which influence the dataflow with distinct characteristics:

- *EventScatterer* components accept a single stream of events as input and fans it out into multiple event streams at the output. Each event is left as is, therefore each output stream only consists a subset of events, corresponding to the fan-out level.
- *EventGatherer* components are the inverse components to EventScatterers. They fan-in multiple input event streams and forward each received event unchanged into to their single output event stream. The EventScatterer/EventGaterer pair is used for load balancing by fanning a data stream out to as many processing streams are required in order to maintain the required event processing rate.
- *EventMerger* components also have multiple input streams and a single output stream. Unlike gatherers they expect one specific part of an event to arrive on every input stream. The descriptors for the input data blocks of these received sub-events are then merged into a combined event with a single event descriptor, which is sent subsequently to the output stream. EventMergers also implement fault tolerance functionality in order to avoid the data flow from blocking in case of one sub-event being lost or delayed due to the reconfiguration of part of the system. The EventMerger maintains lists of incomplete events, which are set aside, while continuing the processing of other complete events.
- *Subscriber-* and *PublisherBridgeHead* components act together in pairs to create a transparent bridge between components on different nodes. The purpose of these so-called *BridgeHead* components is to provide a common interface to publishers and subscribers for processing components, without having explicitly to deal with networking code.

In the Subscriber- and PublisherBridgeHead components network, communication is handled by an abstract class library that provides all required interfaces for the communication. Implementations of these interfaces currently exist for the TCP/IP network protocol and the SCSI[1] API for Dolphin Scalable Coherent Interconnect (SCI)[2] adapters, thereby supporting both the socket based streaming data and the Remote Direct Memory Access (RDMA) paradigm. The bridge components are thus independent of the underlying network functionality, so that only an implementation of the abstract API has to be provided in order to support a new network.

[1]SCSI : Small Computer Systems Interface

[2]Scalable Coherent Interconnect is an IEEE standard that defines the architecture and protocols to support shared-address-space computing over a collection of processors.

### C. The TaskManager

One of the important challenges in the ALICE HLT is the management of the large number of processes distributed in the cluster. It has to be ensured that all processes are started and connected in the correct order. For this purpose the TaskManager [13] has been developed to control and supervise the HLT framework.

The design of the framework supports hierarchical operation with multiple levels of TaskManagers, each controlling subordinate TaskManagers, possibly running at different sites. At the lowest level the slave TaskManagers actually control the framework components. This hierarchical configuration scheme provides several advantages over a flat hierarchy with a single process controlling all components. The TaskManager allows to split up the system into a hierarchy of separate parts, which are easier to handle than a single large configuration. Faults occurring in the system are thus isolated where they occur ensuring that they do not influence the system as a whole. The hierarchical approach also facilitates building a fault-tolerant system, in particular by avoiding single-points-of-failure in the control infrastructure.

Configurations for the TaskManager are stored in XML files. Amongst other items multiple sections of Python code are contained in configuration files. They are executed when a respective event in the TaskManager occurs, e.g. a state change in one of the supervised components. The code thus specifies the actions that are to be taken upon such an event, making the system very flexible. The communication with controlled sub-components is handled via a shared library, specified in the configuration files and loaded at runtime.

## III. The Global Test

### A. Global Test Motivation

The ALICE HLT data transport framework has been designed with quite a specific environment in mind, that of low-latency clustered computing centres, with dedicated bandwidth and processing hardware available locally. The functionality of the framework has indeed been tested in this environment and has been shown to satisfy the constraints imposed on it by the triggering and data taking scenarios of ALICE.

Real-time on-line systems, such as the HLT, usually have high input data rates and stringent processing requirements. The data rates typically exceed the available wide area network bandwidths by up to two orders of magnitude. A further constraint of all real-time systems is the available maximum processing latency per event. However, due to its architecture, the ALICE High Level Trigger does not have any fixed latency requirements. Since the HLT is built by a collaboration with members in Bergen (Norway), Cape Town (South Africa), and Heidelberg (Germany), a globally distributed test of an HLT-like system was considered.

This test served many purposes. First it required a very high degree of flexibility in the framework in order to allow its operation on various different sites with many different installations. It had to work across various firewalls. The control of this system, while running on many independent sites behind firewalls was very complex. On the other hand any unknown cyclic dependency would show as blocking during the global test. Another feature of this setup is the demonstration of a radically different way of distributed processing, being purely based on global, re-routable data streams. The resources at the various centres were under the direct control of the collaboration, and were centrally controlled. The configuration of the test is described below, and was done in Heidelberg, working remotely from Cape Town. This very strict control over the computing resources is somewhat a deviation from the traditional federated grid of computing resources idea and does not present a principle requirement but was rather done in order to simplify the test itself.

Limitations and problems of this approach should be examined in further work if a working system could be obtained. As a first step in this investigation, the test under discussion here did not implement grid authentication or resource management mechanisms. However, these features could also been added in the future, using for example Virtual Private Networks (VPN), or standard lightweight grid middleware. In order to create a full global north-south axis across the globe as well as some east-west expansion two further sites in Tromsø(Norway) and Dubna (Russia) have been included in the setup, in addition to the listed HLT collaboration institutes.

### B. Global Test Configuration

For the global test a configuration was chosen that mimics a part of the ALICE HLT processing. The setup was configured similar to the ALICE's Time Projection Chamber (TPC) [14] and DiMuon Spectrometer [15] process (see figure 2). In order to avoid the bottleneck posed by the available network bandwidth to Cape Town, as detailed below, no actual simulated detector data was transmitted. Instead only mock-up data objects with the same size as that expected during experimental running were sent between the different locations and consequently no real analysis components were used. Therefore mock-up components not performing any processing were running in the setup.

At three of the sites (Bergen, Tromsø, and Dubna) the components were set up to mimic the cluster-finding on data of the TPC sectors. As an example, the setup at one of the sites is shown in Fig. 1, and each patch was processed on a separate node.

The output data produced at these three sites was then sent to Heidelberg, where it was merged. Further mock-up processing steps, which correspond to the tracking in the simulated TPC sector, were performed on two twin processor nodes. On each CPU one mock-up process was active, as shown in Fig. 2. The output produced by these four components was then sent further to Cape Town.

In Cape Town the TPC data stream was merged with the DiMuon data stream. This was generated by another processing chain, fully running in Cape Town as well, which simulated processing raw data from the DiMuon detector [15] from the cluster-finding up to the tracking. The tracked mock-up data was then merged with the received TPC data as the last step in the processing chain. During LHC data taking this component will be the location where the trigger decision is made and/or where the completely reconstructed event data could be written to permanent storage.

The data flow on the different sites is shown in Fig. 3. As can be seen, one characteristic of this test are the relay nodes required by the data flow in order to traverse the firewalls to reach the actual processing nodes. For instance in Heidelberg the cluster access node was not directly accessible from outside the institute so that another relay step was necessary. One advantage of this setup is that inside a cluster all nodes are treated equally and are trusted. Therefore inside the cluster there is only little security functionality required. On these access nodes privileged relay components have been running, which do not touch their input data but only forward it to their output.

Control of the system was provided by a three-level set of TaskManager processes with the top master TaskManager running on the fire wall in Heidelberg. Second-level TaskManagers were setup on the access nodes, communicating both with the top master TaskManager and with the third-level TaskManagers on the cluster nodes themselves, therefore performing bridging functionality. No second-level TaskManager was required on nodes in Dubna, being directly accessible. On each node the local third-level slave TaskManager was used to control the HLT framework processes and ensure their proper operation. The TaskManager hierarchy is shown in Fig. 4.

A second test has been run with a very similar setup, the difference being that instead of mock-up data and mock-up processing components simulated raw data for the ALICE TPC detector and real analysis components for that data have been used. The analysis components used here will also run in the operational HLT. In this set-up the cluster in Cape Town was excluded, as the size of the simulated event data is significantly larger than the mock-up data. Transferring this amount of data over the comparatively slow link to Cape Town was therefore not considered, so that only the clusters in Tromsø, Bergen, Dubna, and Heidelberg were used.

### C. Global Test Results

All tests discussed below demonstrate the successful operation of the HLT on-line framework in a grid-like environment. For the first setup an initial test was started with the event rate limited explicitly to 10 Hz, in order not to over-stress the network link to Cape Town. This test had been running overnight unattended for more than 15 hours. During this time more than 500,000 events have been passed through the processing chain before the test was stopped by the operators. In a second test with the same configuration the limit to 10 Hz was deactivated and the chain was allowed to run at the maximum achievable rate. During the test's runtime of about two hours a maximum rate of 15 Hz was reached. Note that the speed of light in vacuum corresponds to 40 ms over the given distance of 12000 km. One round trip delay results in a theoretical 12 Hz limit. This test therefore demonstrates that there is no global flow control our round trip delay anywhere in the system. All communications are performed pipelined and point to point. The number of events, being processed simultaneously and the corresponding latency and memory requirements in the system scales with the system size. This system effectively implements a more than 10000 km long physical pipeline, operating at the limits of the slow network link to Africa. A third test was then run using the second configuration, excluding the Cape Town site but using realistic simulated ALICE TPC detector data and real analysis components on the four remaining sites. This configuration has been running at 3.3 Hz for about 100 min., therefore processing approximately 20000 events. In order to start the overall system, all tasks on all nodes had to be started in a co-ordinated way. Any error in this process would require to restart the entire procedure. The built-in fault recovery mechanisms, however, allowed to remedy many errors locally, without completely restarting the chain. Without this feature the tests would not have been successful during the short time available for the tests. In the long run the repair mechanisms - restarting processes and reconnecting them to the framework - will be done automatically by appropriate intelligent daemons, currently being developed.

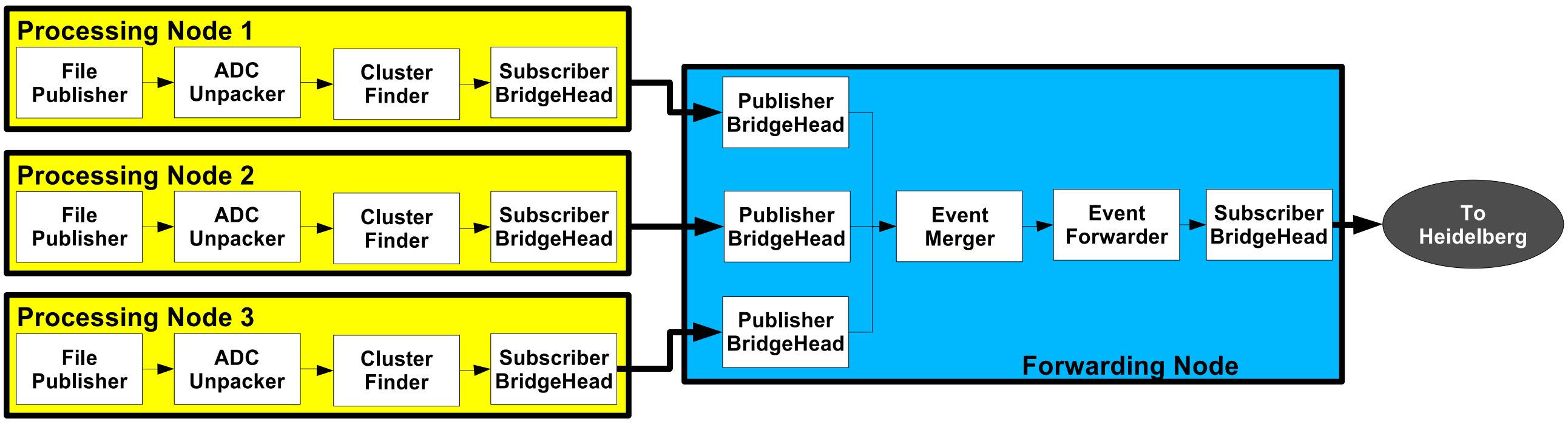

Fig. 1. Setup of the cluster-finding processing steps at the Bergen site.

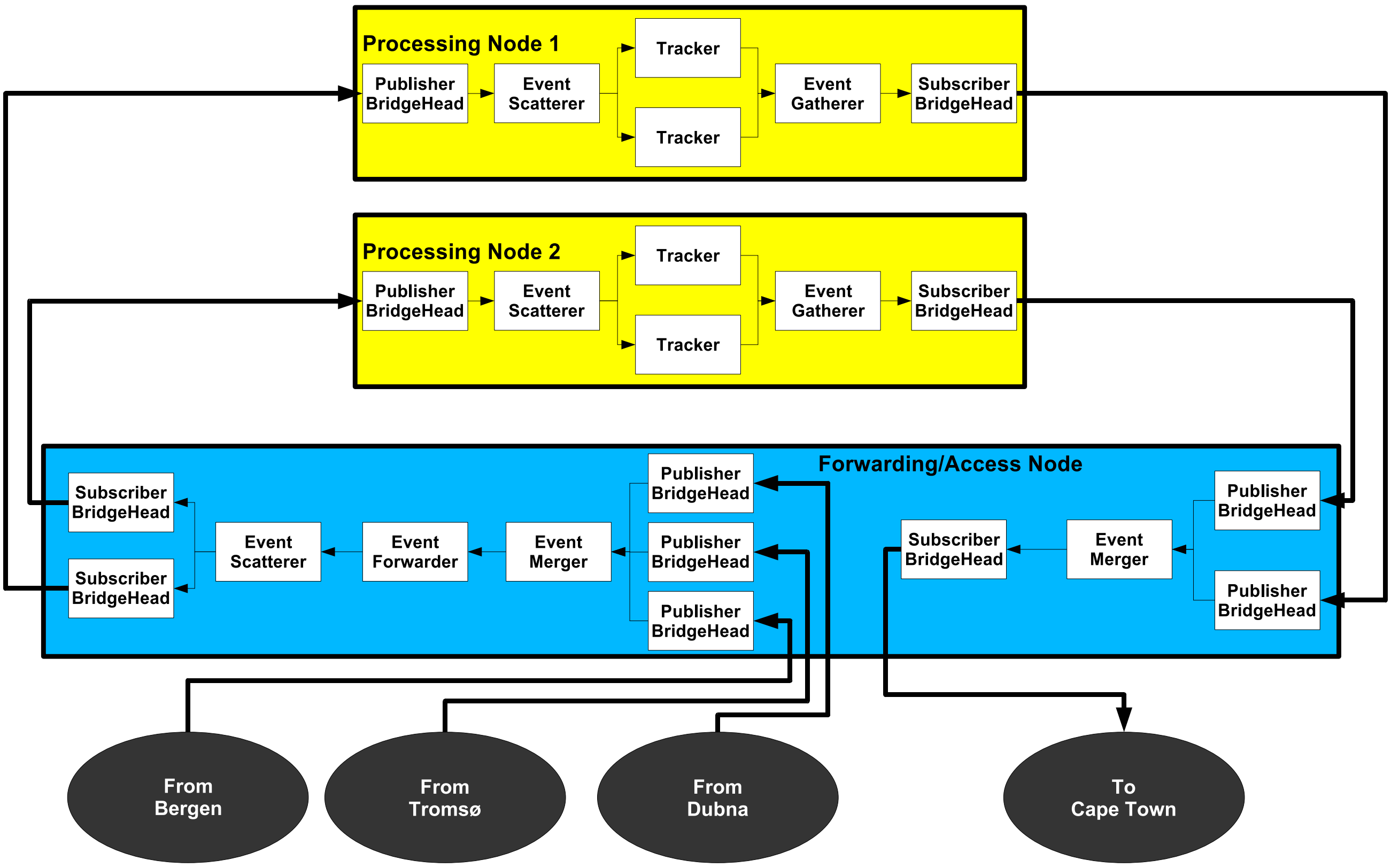

Fig. 2. Setup of the tracking processing steps at the Heidelberg site.

In all test cases the bottleneck was clearly network bandwidth between the sites, for the first two tests in particular the one to Cape Town. This was mainly due to the fact that the test had been run on normal working days using the sites' normal internet connections as links. The links were therefore busy with the site's basic traffic, leaving only a limited amount of bandwidth available for the test. Using dedicated links the achievable rates should be higher, correspondingly with the increase in bandwidth.

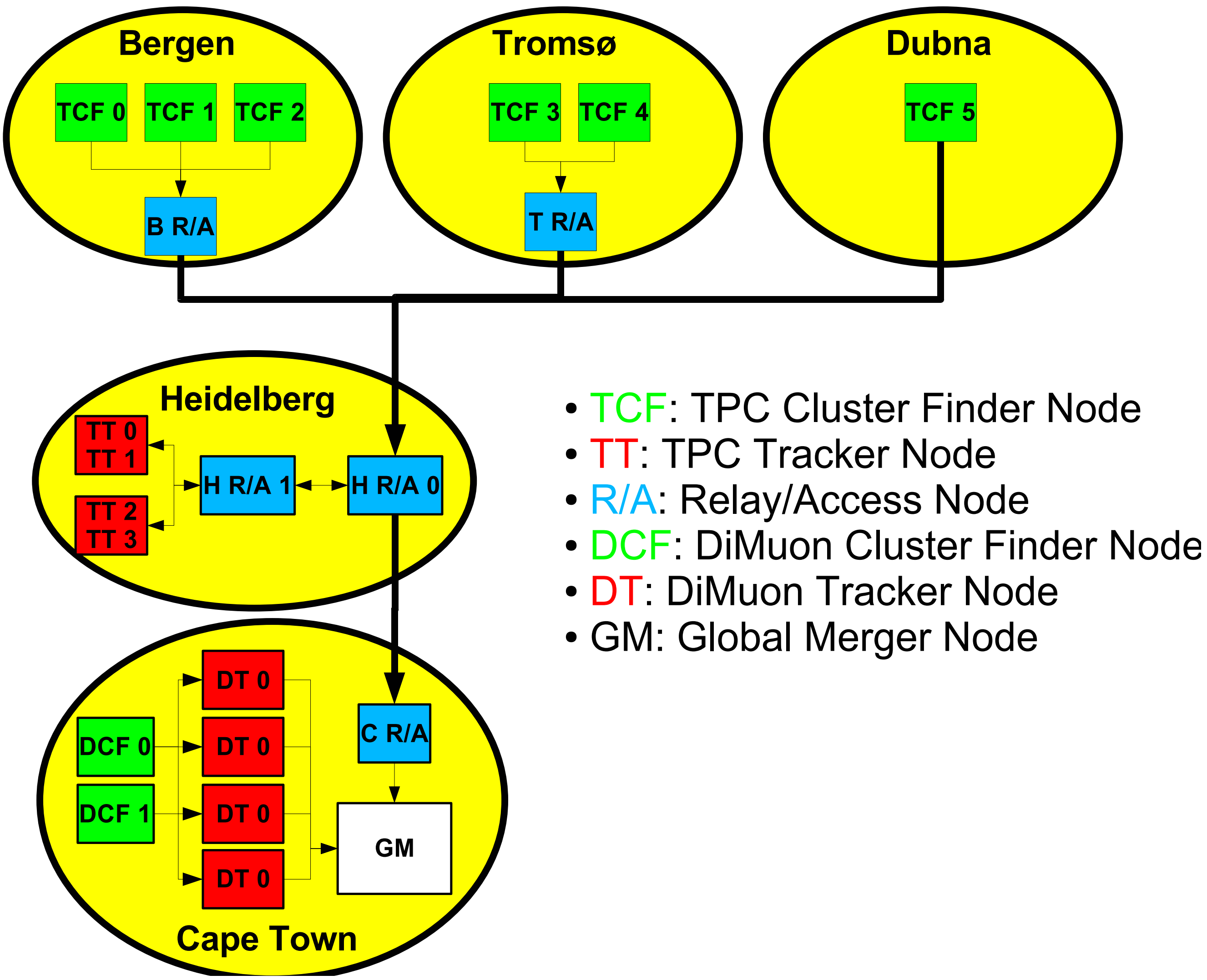

Fig. 3. The global setup with all involved sites and nodes. Components on each node are not shown.

The ALICE HLT is targeted to run at 200 Hz for TPC events and up to 1 kHz for DiMuon only events. The required rates here are about a factor 30 too low. However, the goal was not to prove the feasibility to run the HLT application as grid application under current conditions, as this would require a 25 GB/sec bandwidth into public networks, which would be prohibitively expensive. The goal was to demonstrate that such a system is possible in the first place and secondly is working. It can be adapted to any application with a somewhat lower networking requirement.

## IV. Summary and Conclusions

As an overall result it can be stated that the approach of using the ALICE High Level Trigger data transport framework for a grid-like system was successful. The framework, which was designed for use in a single cluster configuration, has functioned as desired in a globally distributed system. Performance of the system was restricted only by the limited amount of bandwidth available for the tests on the normal Internet connections of the involved sites. Therefore, the use of distributed grid-like online systems has been shown to be feasible in principle. The fundamental requirements are primarily defined by the bandwidth requirements of the system involved. The latencies incurred due the large distances do not matter in this application and only require some additional on-line storage space in order to maintain the deeper pipelines. The HLT on-line system will implement queues, supporting thousands of events or several seconds of running time. Therefore the latency incurred even in a system of the presented scale will not make any difference.

We have demonstrated the successful operation of a data driven, distributed processing paradigm, operating on global scale. There are no particular latency constraints. The transaction rate of the system depends only available network bandwidth and data sizes, while the depth of the pipeline is irrelevant. Therefore the ALICE HLT software framework can in principle be operated over arbitrarily large distances, given a TCP/IP connection. However, making a complex, distributed system operate at all at the presented scale is an achievement, documenting how powerful the communication framework is. Depending on the data rate requirements and available networks, which could be dedicated, this software infrastructure is suitable for any data driven, distributed processing, in particular for applications where the data is being acquired at various, distributed sites, therefore lending itself to a data driven, distributed processing system.

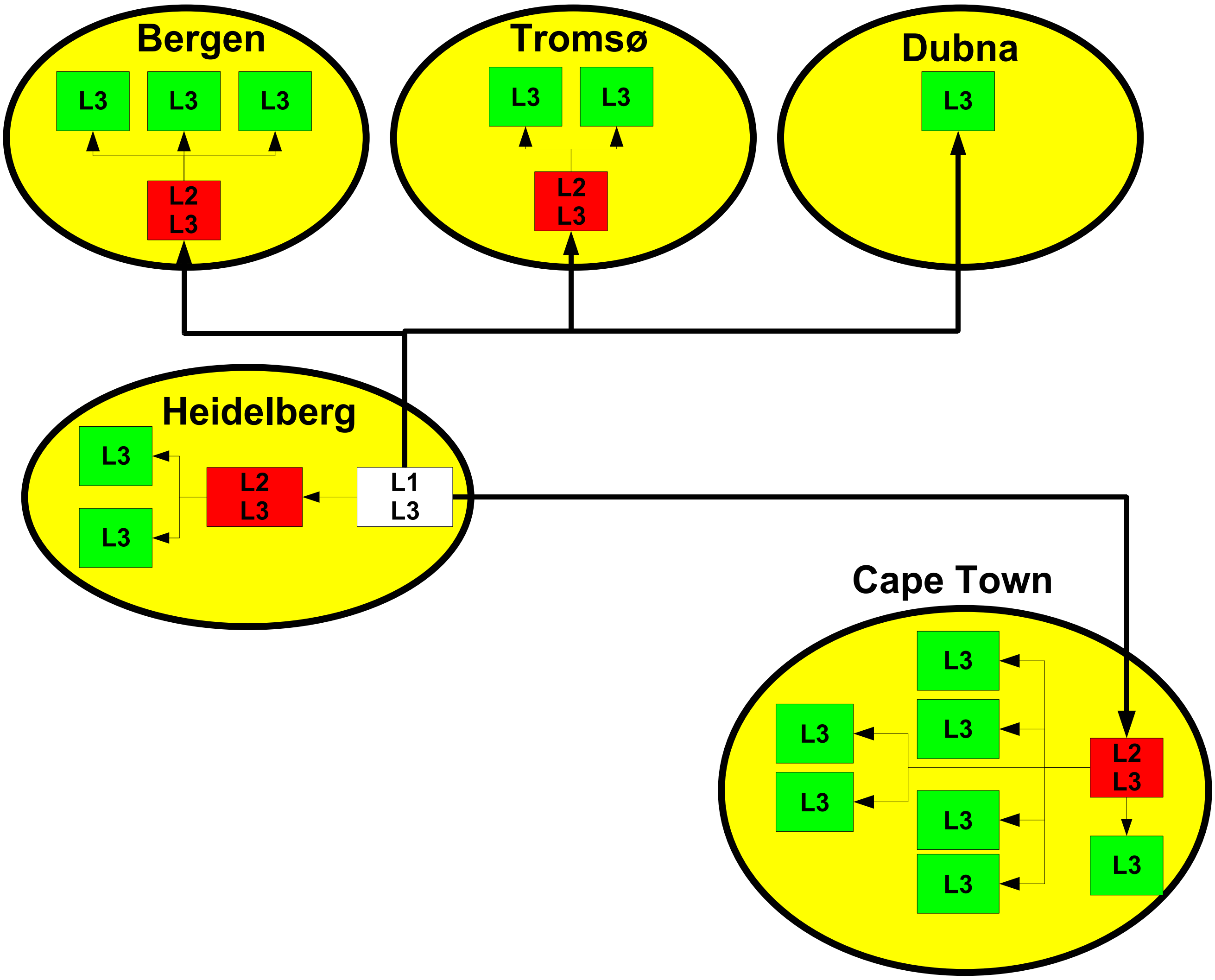

Fig. 4. The hierarchy of the TaskManager control processes. Each L3 TaskManager slave is controlled by the closest L2 intermediate TaskManager or in the case of directly accessible nodes by the L1 master TaskManager.